# Optimization of Plasma Plume Characteristics Based on Multi-anode Coaxial Ablative Pulsed Plasma Thruster

Weisheng Cui, Wenzheng Liu*, Yongjie Gao

*School of Electrical Engineering, Beijing Jiaotong University, 100044 Beijing, People's Republic of China*

**Abstract**
A special surface discharge is proposed based on the multi-anode electrode geometry. Instead of the traditional surface flashover of creepage on the insulator surface between the electrodes, a surface discharge with one of the electrodes being placed far away from the insulator is achieved in this paper. The unique electric field distribution due to the multi-anode electrode geometry has a significant influence on the discharge process of the surface discharge. It changes the generation and propagation process of the plasma, forming a plasma plume contributes to the propulsion performance of the thruster. Through theoretical analysis of the obtained plume data, it is indicated that the ablative pulsed plasma thruster based on multi-anode electrode geometry (short for multi-anode APPT) promotes the internal pressure of the plasma jet during its propagation and significantly increases the density and energy of charged particles. The discharge phenomena manifest that the multi-anode APPT and the helix-coil multi-anode APPT effectively increase the intensity of the plasma plume. Through electron density spatial distribution measurement, it has been found that the helix-coil multi-anode APPT increases the density of plasma in the axial direction to more than 4 times of the conventional coaxial APPT and reduces the electron density in other directions. In the propulsion test, it has been demonstrated that the multi-anode APPT and the helix-coil multi-anode APPT have better performance in terms of the impulse bit and the thrust-to-power ratio. In addition, it is also identified that the pinch effect will be enhanced with the increase of discharge power and the propulsion performance is promoted more distinctly. The multi-anode APPT and the helix-coil multi-anode APPT have been proved to have potential application advantage in the field of micro-satellite propulsion.

## I. INTRODUCTION

The existing Ablative Pulsed Plasma Thruster (short for APPT) is one of the promising electric propulsion devices because of its simple structure, small volume and high reliability [1-6]. In recent years, the rapid development of micro-satellites has put forward higher requirements on the propulsion performance of APPT [7-9]. While, the radial diffusion of charged particles exists in the plasma plume, which will affect the propulsion performance of APPT. Actually, beam divergence has become one of the most negative characteristics of APPT. Therefore, the optimization of plasma plume to increase its density is critical to the improvement of the propulsion performance of APPT.

Z-pinch is an important method of plasma constriction in the study of plasma characteristics [10-14]. In 1959, Haines[15] described the constraint characteristics of an axial current sheet on the plasma column. Keidar [12] built a model that included Joule heating of the plasma, heat transfer to the Teflon and Teflon ablation, and concluded that the pinch effect and the thrust-to-power ratio increase with the pulse energy. Markusic [14] studied the pinch effect of current sheet distributions in three different APPT structures and found that the thruster with a spike anode could increase the efficiency and specific impulse. However, all of the confinement are limited within the thruster, and the issue of plasma diffusion after the plasma leaves the nozzle has not been considered.

The application of an external magnetic field by additional circuit or permanent magnet is another effective method to optimize the plasma plume [16-22]. Hu [23] investigated the performance of a thruster with different outer diameters of magnet rings, and found that the magnetic field strength will affect the utilization of propellant. Takahashi[24] measured the axial magnetic field induced by the plasma flow in a divergent magnetic nozzle and detected a decrease in the axial magnetic field strength near the source exit, whereas an increase at the downstream side of the magnetic nozzle. Nevertheless, the use of circuit or permanent magnet to form magnetic fields will increase the size, quality or the complexity, which is not conducive to its application in APPT.

In this paper, we achieve a surface discharge on the condition that one of the electrodes is far away from the insulator and the creepage path is truncated. We apply this type of surface discharge to the coaxial APPT to improve its performance. This discharge method is able to realize a complete plasma plume confinement in the whole propagation process of plasma. It solves the problem that the conventional Z-pinch cannot effectively act on the plasma leaving the nozzle and the increase of the quality or complexity caused by the application of external magnetic field. This paper is organized as follows: the experimental system is introduced in the second part. The APPT based on the multi-anode electrode geometry (short for multi-anode APPT) is proposed along with its comparison with the conventional coaxial APPT (short for conventional APPT) in the third part. This part elaborates the special discharge process of the multi-anode electrode geometry and verifies the pinch effect of the multi-anode APPT. In the fourth part, a helix-coil multi-anode APPT and its characteristics are introduced. The fifth part validates the performance improvements of the multi-anode APPT and the helix-coil multi-anode APPT and it is followed by the conclusion and prospect.

## II. EXPERIMENTAL SETUP

The experimental system mainly includes a pulsed power supply circuit, a vacuum discharge device and diagnostic systems, which is shown in Fig. 1. The specific discharge procedure is as follows: the energy storage capacitor C (0.1 μF) is firstly charged to a high voltage, and then the spherical gap SG is connected by a pulse signal. At this moment, the voltage of C will be applied between the cathode and the anode through the current-limiting resistor R (27 Ω) and the inductance L (220 μH), so that the discharge is produced due to surface flashover. The L is designed to prolong the duration time of arc current and promote the generation of plasma. A diode D in series with the circuit is used to prevent the circuit current from flowing back and avoid the occurrence of current oscillation, thus increasing the discharge stability of the APPT. The output of this pulsed power supply is a negative voltage with a maximum amplitude of 30 kV. The discharges in this experiment are carried out under high vacuum environment and the pressure is maintained at $10^{-4}$ Pa. During the discharge, the interelectrode voltage (voltage between point A and the ground) is measured by a high voltage probe (TEK-P6015A); the currents flowing through the cathode and the anode (short for cathode current and anode current) are obtained by two Rogowski coils. The photograph of the discharge is captured by a camera and the plasma electron density is measured by an improved Langmuir probe method [25-28].

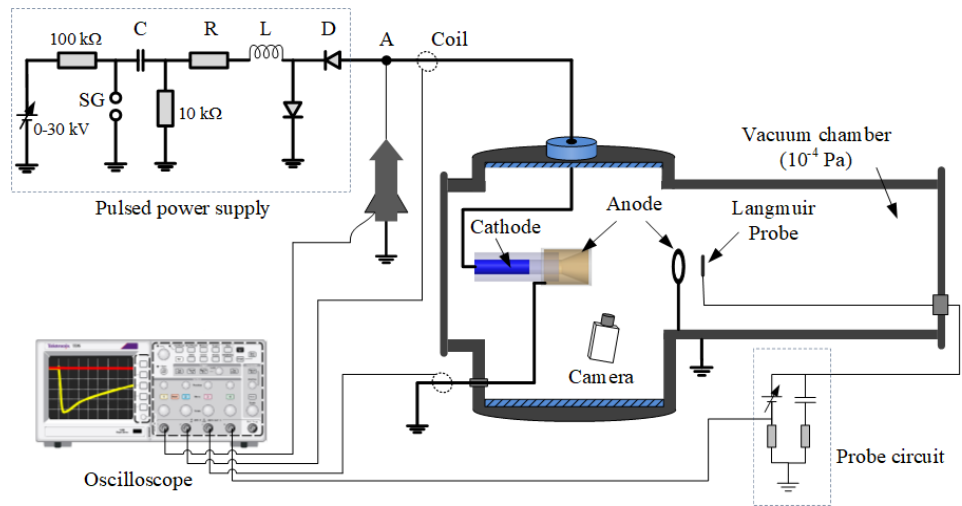

FIG. 1. Schematic diagram of the experimental system.

## III. THE MULTI-ANODE APPT BASED ON Z-PINCH

### A. Propagation characteristics of the plasma plumes for multi-anode APPT

The APPTs in this paper are based on coaxial electrode geometry. In the conventional coaxial APPT studied in this research, a solid polytetrafluoroethylene (PTFE) tube is used as the propellant. Its inner diameter and outer diameter are 5 mm and 10 mm respectively. The cathode and the anode are made from stainless steel material: the cathode is cylindrical and is placed in the PTFE tube; the anode is in the shape of a trumpet nozzle and is fixed to one end of the PTFE tube. The axial distance between the cathode and the outlet of PTFE tube is 10 mm. The triggering mechanism of cathode triple junction (short for CTJ) [29-33] is used to initiate the surface discharge between the cathode and the anode.

It is commonly accepted that the surface discharge usually requires both the cathode and the anode to contact with the insulator. The secondary electrons and charged particles will form an arc path along the insulator surface between the electrodes. The type of surface discharge of one electrode contacting with the insulator and the other electrode being far away from the insulator has not been proposed yet. This situation is concluded to be due to the following reasons: the separation of the electrode and the insulator results in the truncation of the creepage path of surface flashover; the weakening of the electric field intensity in CTJ due to the long distance between the electrodes makes it hard to emit initial electrons; it is difficult to generate enough plasma to connect the cathode and the anode before the formation of arc current.

In our previous research, we proved that the generation of plasma will continue provided that the strong electric field in the CTJ is sustained. Based on this theory, a method of using specially designed multi-anode electrode geometry to realize a long distance-noncontact surface discharge is proposed in this paper. The multi-anode electrode geometry is designed on the basis of conventional coaxial electrode geometry. A anode next to the cathode is completely insulated and is the first anode (short for insulated-anode). It is able to maintain the electric field intensity on the CTJ before the formation of arc current. A second anode (short for remote-anode) is in the shape of a ring and is arranged 100 mm axial direction away from the insulated-anode nozzle. The remote-anode has an inner diameter of 5 mm and an outer diameter of 15 mm. It is exposed to complete the discharge.

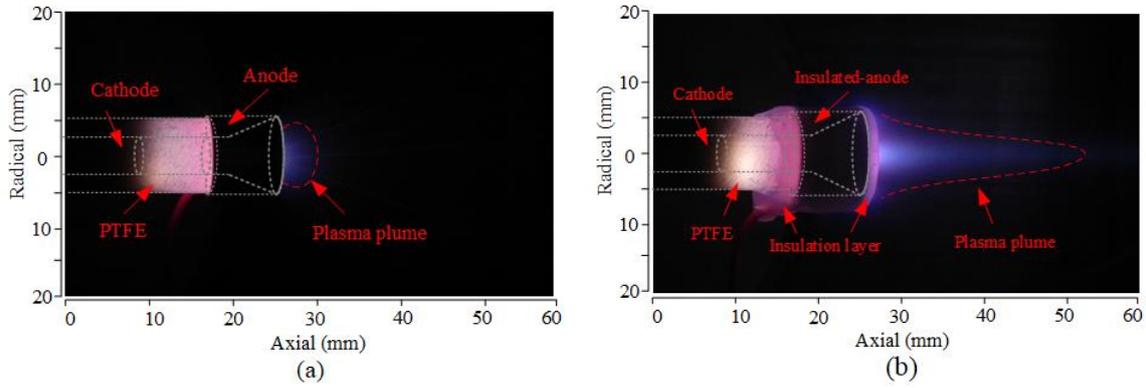

FIG. 2. The plasma plume of the conventional APPT (a) and the multi-anode APPT (b) (the second remote-anode is not shown in the graph because of its long distance with the APPT).

The photographs of the plasma plume for the two APPTs with the same power supply parameters are captured and shown in Fig. 2. The schematic diagram of plume boundary has been indicated in the graphs. It has been observed from Fig. 2 that the plasma plume of the conventional APPT is weak and divergent outside the anode nozzle, while the plasma plume of the multi-anode APPT is enhanced obviously and exhibits a trend of shrinking with the increase of the distance with the nozzle. It is inferred that the plasma plume of the conventional APPT diffuses after it leaves the nozzle due to the lack of constraint. The plasma of the multi-anode APPT is completely pinched in the whole propagation process, so the plasma exhibits a convergence and the intensity is increased. The operation principles of the two APPTs are shown in Fig. 3.

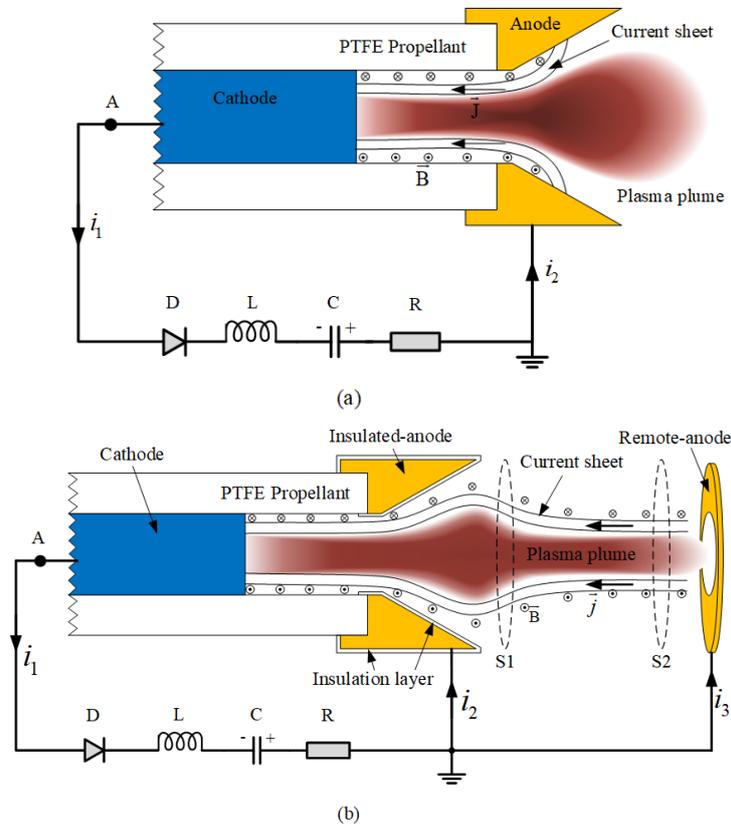

FIG. 3. Schematic diagrams demonstrating the operation principle of the conventional coaxial APPT (a) and the multi-anode APPT (b)

In the coaxial APPT, the initial current distribution is usually in the form of a cylindrical current sheet along the face of the insulator. The current sheet forms at the cavity's maximum radius because it is the path of least impedance available to the circuit [14]. As the current increases, an azimuthal magnetic field is formed on the outside of the current sheet; this magnetic field interacts with the current sheet to create a radially inward directed electromagnetic self-force, which is the principle of Z-pinch. However, as shown in Fig. 3(a), the current sheet exists only between the cathode and the anode in conventional APPT, so that the Z-pinch only works on the plasma inside the PTFE tube. The diffusion of the plasma flowing out of the anode nozzle still exists and affects the performance of the thruster. It has been seen from Fig. 3(b) that the current sheet can be formed between the cathode and the remote-anode in multi-anode electrode geometry, which produces a Z-pinch not only on the plasma within the PTFE tube, but also on the plasma out of the nozzle.

## B. Discharge characteristics of the multi-anode electrode geometry

The cathode and the first anode in multi-anode electrode geometry have the same parameters with the conventional electrode geometry (except for the insulation layer). Therefore, the electric field intensity in the CTJ for the two electrode geometries is about the same. The initiation mechanism of the surface discharge for the multi-anode electrode geometry is similar to that of the traditional surface flashover. The electrons emitted from the CTJ bombard the PTFE and the desorbed gas molecules to generate the initial plasma on the surface of insulator, which is formed by secondary electrons[34,35] and charged particles ($C^+$, $F^+$, etc.)[36-38]. The cathode and the anode will be connected directly along the PTFE surface in traditional surface flashover. However, the first anode contacting with the PTFE surface is insulated and the surface leading to the second anode is truncated in the multi-anode electrode geometry. As a result, the cathode is not able to be connected to both anodes through surface flashover at this moment.

In order to investigate the discharge process of multi-anode electrode geometry, the ANSYS Maxwell 3D software was used to analyze the electric field distribution of the electrode geometry. When a voltage of 10 kV is applied between the cathode and the anodes (the insulated-anode an the remote-anode), the distribution of electric field vector of the multi-anode electrode geometry is shown in Fig. 4.

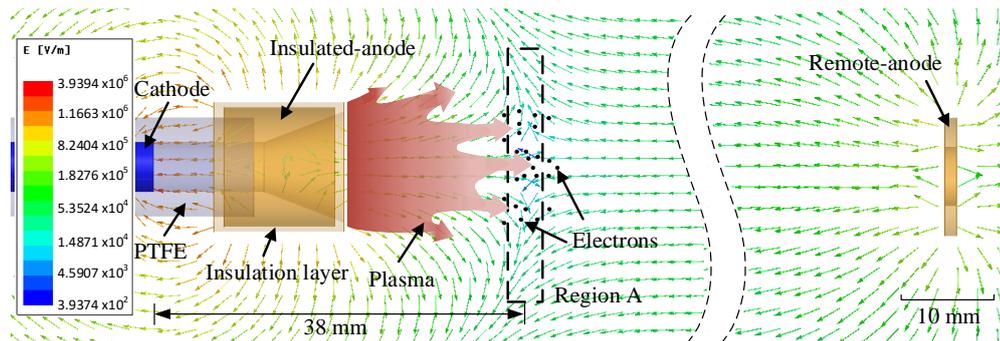

FIG. 4. The distribution of electric field vector of the multi-anode electrode geometry. The schematic diagram of the plasma and the extracted electrons are depicted in the graph.

As shown in Fig. 4, there is a region (called region A) between the insulated-anode and the remote-anode, where the electric field vector has opposite directions. Through further simulation, it has been found that the region A remains unchanged before the arrival of the plasma; and it moves together with the front edge of plasma when the plasma arrives at its position and propagates towards the remote-anode. It has been proved that the existence of region A has an important influence on the initial process of the surface discharge .

It is generally accepted that the directional drift of electrons is mainly determined by the electric field and the propagation of the plasma is affected by the ions (here, referring to $C^+$, $F^+$, etc.). Therefore, the diffusion of electrons towards the remote-anode is inhibited by the electric field between the insulated-anode and the region A, leading to a sustained high interelectrode voltage. The maintenance of the high interelectrode voltage guarantees the continued generation of plasma on the surface of PTFE, which contributes to the propagation of plasma towards the remote-anode. When the plasma arrives at the region A during the propagation, some electrons in the front edge of the plasma will be extracted and reach the remote-anode instantly under the influence of the electric field between the region A and the remote-anode. When the plasma propagates to the remote-anode, the arc path is able to be formed between the cathode and the remote-anode.

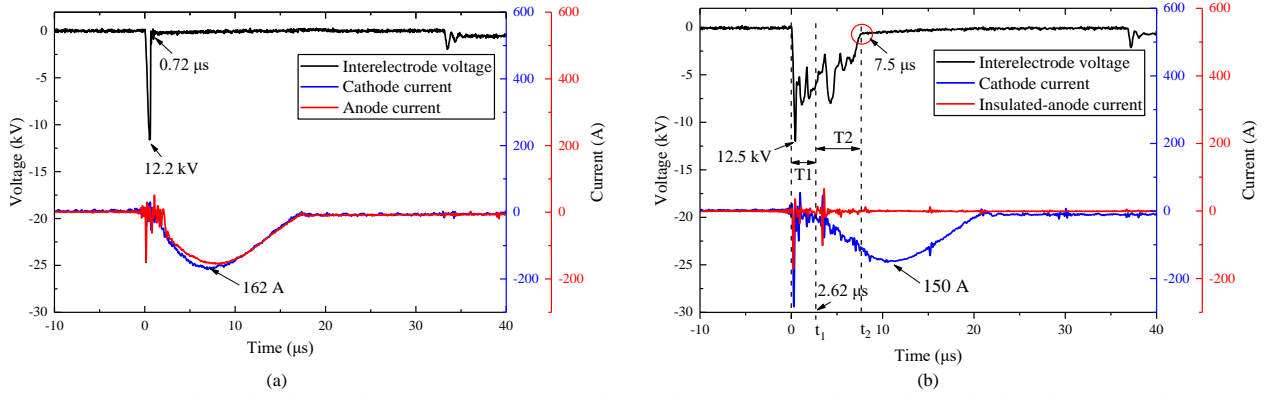

FIG. 5. Interelectrode voltage and current of the conventional electrode geometry (a) and the multi-anode electrode geometry (b).

A series of experiments have been carried out and the typical waveforms of the interelectrode voltage and the discharge current of the two electrode geometries with the same discharge parameters are obtained and shown in Fig. 5. It has been seen from Fig. 5(a) that the surface flashover of the conventional electrode geometry is achieved instantaneously and the interelectrode voltage drops to arc voltage within 0.72 μs (the arc voltage is below 200 V and not displayed in the graph). Afterwards, the cathode current and the anode current remain approximately the same in the discharge process. According to the Kirchhoff current law, the current flowing through the plasma plume is tiny. As a result, the plasma is not confined in its propagation outside the nozzle.

In contrast, the interelectrode voltage and current waveform of the multi-anode electrode geometry exhibits a completely different discharge characteristic. Fig. 5(b) shows that the current flowing through the insulated-anode (short for insulated-anode current) is basically zero during the discharge. The interelectrode voltage and the cathode current have different characteristics in different time intervals. The plasma propagation velocity in this APPT is referred to, which is in the range of 12.6-16.4 km/s according to the previous study [26]. According to the distance between the cathode and the region A (about 38 mm), assuming the velocity of the plasma is constant during the propagation, the time needed for the plasma to arrive at region A is 2.32-3.02 μs. It has been seen from Fig. 5(b) that the interelectrode voltage and the cathode current do not exhibit obvious change in the time interval of 0-2.62 μs (T1), which corresponds to the calculation. The cathode current grows gradually and the interelectrode voltage starts to drop to the arc voltage in the second time interval of 2.62-7.5 μs (T2). In this period, the plasma arrives at the region A and the region A moves with the front edge of the plasma. An increasing number of electrons are extracted in the region A and reach the remote-anode, leading to the change of interelectrode voltage and the cathode current. When the plasma reaches the remote-anode, the arc is formed and the interelectrode voltage drops to the arc voltage.

The time needed for the plasma to propagate to the remote-anode is calculated to be 6.7-8.8 μs according to the distance between the cathode and the remote-anode. It is consistent with the time when the interelectrode voltage drops to the arc voltage. The distance between the remote-anode and the cathode is altered in subsequent experiments to further verify this hypothesis. The time when the interelectrode voltage drops to the arc voltage is recorded for each configuration. It drops to the calculated time zone based on the distance and the plasma propagation velocity each time. As a result, the analysis has been confirmed.

### C. Theoretical analysis of Z-pinch on the plasma plume

In order to investigate the Z-pinch on the plasma plume of the multi-anode APPT, we established a theoretical model of the plasma internal pressure relative to the axial distance. The plasma, as a fluid containing charged particles, can be described using the theory of magnetohydrodynamics. The equations of magnetohydrodynamics and electrodynamics are written as

$$\rho \frac{d\vec{u}}{dt} = -\nabla p + \vec{F} \quad (1)$$

$$\vec{F} = \frac{1}{\mu_0}(\nabla \times \vec{B}) \times \vec{B} \quad (2)$$

Where $\mu_0$ is the vacuum permeability; $\rho$ and $\Delta p$ are the mass density and pressure of the fluid, respectively; $\vec{u}$ is the macroscopic velocity of the fluid and $\vec{F}$ is the force acting on the unit volume of the fluid. According to equation (1) and equation (2), a cylindrical coordinate system (r, θ, Z) is built based on the experimental electrode geometry. Assuming the current remains constant and the plasma column is axisymmetric and Z independent, the relationship between current density $\vec{j}$, magnetic field $B_\theta$ and pressure $dp$ can be obtained from the static equilibrium equation of magnetic fluid.

$$\frac{dp}{dr} = -\frac{B_\theta}{\mu_0}\frac{dB_\theta}{dr} \qquad (3)$$

$$\vec{j} = \frac{1}{\mu_0}\left(0, 0, \frac{dB_\theta}{dr}\right) \qquad (4)$$

After that, we assume the current sheet is cylindrical and the current density is distributed evenly. The current density distribution function of the plasma jet cross-section is expressed as

$$j_z = \begin{cases} 0, & r < r_0 \\ j_0, & r_0 < r < R_0 \\ 0, & r > R_0 \end{cases} \qquad (5)$$

where $R_0$ and $r_0$ are the outer radius and the inner radius of the current sheet, respectively. By substituting equation (5) into equation (3) and equation (4), we are able to obtain the relationship between the plasma internal pressure $p_\theta$ and radial displacement $r$,

$$p_\theta = \begin{cases} \dfrac{\mu_0 j_0^2}{4}\left(\dfrac{R_0^2}{2} + \dfrac{r_0^4}{2R_0^2} - r_0^2\right), & r \leq r_0 \\ \dfrac{\mu_0 j_0^2}{4}\left(\dfrac{R_0^2}{2} + \dfrac{r_0^4}{2R_0^2} - \dfrac{r^2}{2} - \dfrac{r_0^4}{2r^2}\right), & r_0 < r \leq R_0 \\ 0, & r > R_0 \end{cases} \qquad (6)$$

Assuming that the current sheet thickness ($R_0 - r_0$) remains unchanged during the constriction of the plasma jet, the cross-sections in different positions of plasma jet are analyzed (S1 and S2 in Fig. 3(b)). The distribution of cross-section parameters of plasma jets caused by Z-pinch is shown in Fig. 6(a). From parameters variation from the section S1 to the S2, it has been seen that when the plasma jet is reduced to a smaller radius, the current sheet density $j_0$ will increase, so is the internal plasma pressure.

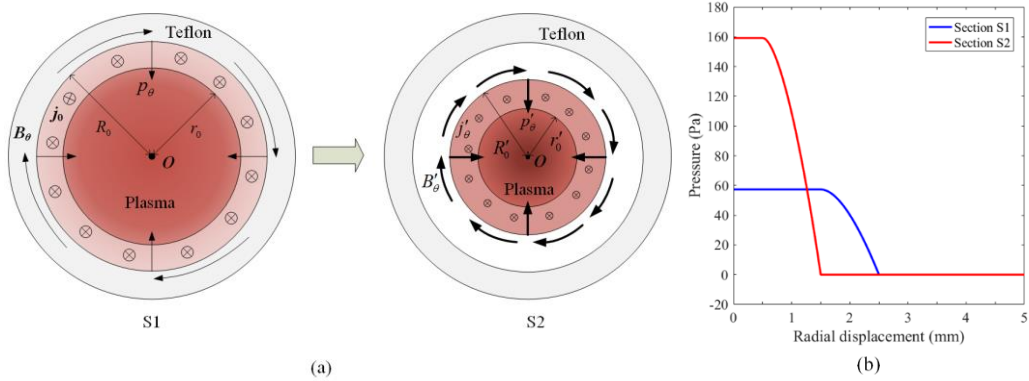

FIG. 6. The change of cross-section of the plasma jet (the deepening of the color indicates the increase of charged particle density) (a) and the radial pressure distribution of the cross-section of plasma jet (b)

Assuming the current is 150 A; $R_0$ and $R_0'$ are 2.5 mm and 1.5 mm respectively; the current sheet thickness is 1 mm; the variation curves of $p_\theta$ for $r$ in the two sections S1 and S2 are shown in Fig. 6(b). It has been seen from the curves that the pressure inside the current sheet increases gradually from the external to the internal, and the pressure in the internal plasma region remains the same as its current density is zero. At the same time, it is calculated that when the outer radius of current sheet decreases to its 60%, the pressure inside the plasma increases nearly 3 times, showing a significant upward trend.

To obtain the relationship between the density of the plasma plume and its axial displacement, the Matlab software is used to analyze the light intensity of the plasma plume and obtain the data of plasma plume cross-section radius relative to the axial displacement. The Origin software is used to curve-fit the data to obtain the plume boundary function. The internal pressure of the plasma at different axial position is calculated with equation (6). The three-dimensional diagram of the plasma jet and its inner pressure is obtained and depicted in Fig. 7. The Z axis in Fig. 7 is the axial displacement of the plasma jet leaving the nozzle, and the color bar on the right represents the pressure inside of the plasma.

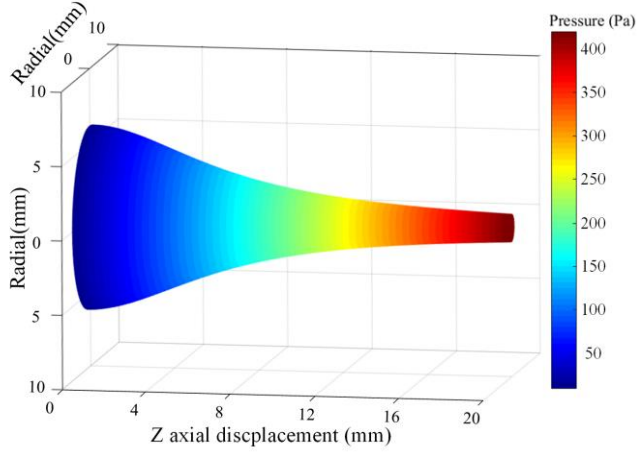
FIG. 7. The pressure distribution of the experimental plasma plume for the multi-anode APPT

It has been seen from Fig. 7 that the plasma is constricted due to Z-pinch when it flows out of the cavity. And the reduction rate of the section radius of plasma jet slows down with the increase of axial displacement. The reason for this result is considered to be that the internal pressure (thermodynamic pressure) increases with the decrease of the plasma jet cross-section radius. This pressure is opposite to the magnetic force formed by Z-pinch and hinders the decrease of jet radius. According to the law of thermodynamics, $p = k\left(n_i T_i + n_e T_e\right)$, the density and temperature of charged particles inside of the plasma will be promoted when the pressure $p$ increases. Therefore, the pinch effect produced by multi-anode electrode geometry on one hand reduces the diffusion of charged particles, on the other hand increases the density and energy of the charged particles. The increase of particle density and energy in plasma is crucial to the performance of thruster.

### D. Experimental verification of Z-pinch on the plasma plume

The spatial distributions of plasma electron density for the two APPTs are measured to verify the experimental effect of Z-pinch. Because the plasma is approximately neutral, the distribution of electron density is able to reflect the distribution state of the plasma. The specific measurement method is shown in Fig. 8(a), where the distance L (120 mm) between the probe and the APPT remains the same at different angles. The spatial distribution of electron density is presented in polar coordinate system, where the angle of 0° indicates the axial direction and the radial displacement represents the value of electron density. The electron density spatial distribution of plasma plumes for the two APPTs are shown in Fig. 8(b) and (c).

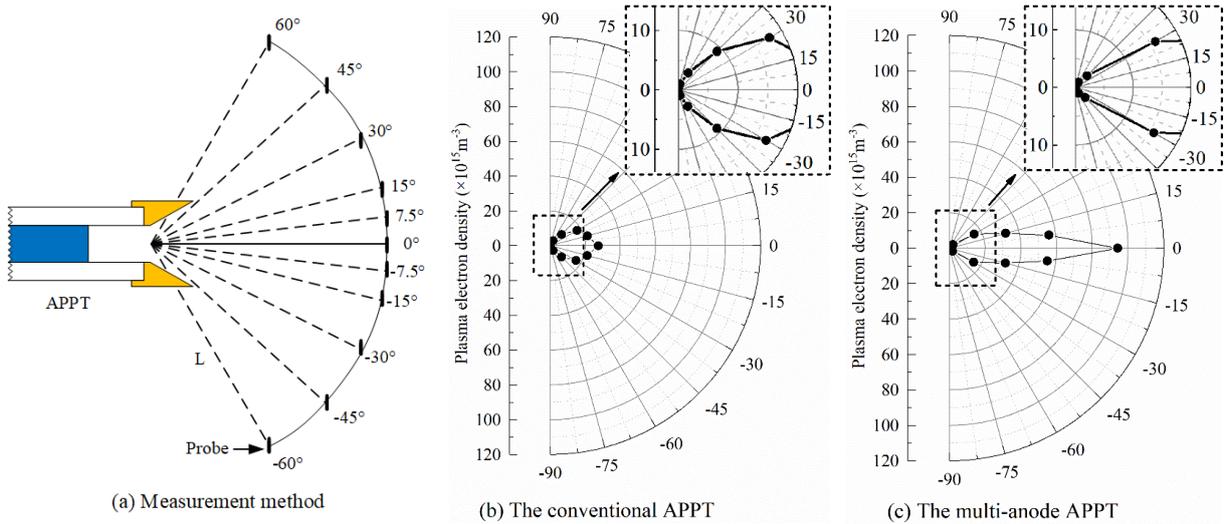
FIG. 8. the measurement method of the electron density spatial distribution (a), and the electron density distributions of plasma plume for the conventional APPT (b) and the multi-anode APPT (c).

As can be seen from Fig. 8, both the conventional APPT and the multi-anode APPT exhibit an electron density distribution of large in axial (0°) and smaller with the increase of the angle. However, the electron density spatial distribution of the multi-anode APPT has been significantly optimized. The electron density at 0° for multi-anode APPT increases to 9.5 $\times 10^{16}$ m$^{-3}$, about four times of the conventional APPT. The decreases of electron density in other angles are also obvious:

the electron densities for the conventional APPT at 15° and 30° are still 80.2% and 62% of the density at 0°; in contrast, the electron density for the multi-anode APPT becomes 34.4% and 16.7%, showing a dramatically decrease. When the angle is greater than 30°, the electron density for the conventional APPT decreases relatively gently, while the electron density for the multi-anode APPT drops rapidly to below $0.5 \times 10^{16}$ m$^{-3}$ (5%). The reduction of the electron density in radial directions contributes to the advance of plasma plume utilization efficiency.

The spatial distribution of the electron density proves that the plasma plume of the conventional APPT disperses in radial directions because of the internal pressure, leading to a poor directionality. In contrast, the plasma plume of the multi-anode APPT is subjected to the pinch force and has a distinct directionality. The increase of plasma density in axial direction will bring some benefits to the performance of the thruster.

## IV. THE OPTIMIZATION OF MULTI-ANODE APPT BASED ON θ-PINCH

### A. The helix-coil multi-anode APPT

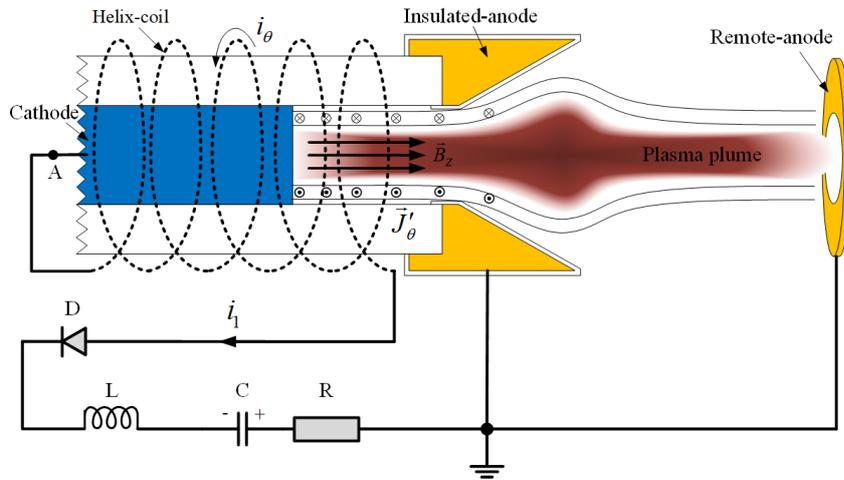

FIG. 9. A schematic diagram demonstrating the operation principle of the helix-coil multi-anode APPT

An approach of producing θ-pinch on the basis of multi-anode electrode geometry is proposed. It uses a helix-coil to reasonably arrange the circuit current path, so that it is called the helix-coil multi-anode APPT. As shown in Fig. 9, a helix-coil is nested outside of the PTFE tube. The left side of the coil is connected to the cathode, and the right side of the coil is connected to the negative output of the pulsed power supply. When the current $i_\theta$ flows through the helix-coil, an induction current $\vec{j}'_\theta$ in θ direction will be induced on the surface of the plasma column. Meanwhile, the magnetic field $\vec{B}_z$ generated by $i_\theta$ is in the Z direction. The $\vec{j}'_\theta \times \vec{B}_z$ will produce a radial pinch force on the surface of the plasma column, which compresses the plasma towards the center. The peculiarity of this kind of θ-pinch relative to the external magnetic field method is that it is produced by its own discharge current. Hence, it exists automatically when the discharge occurs and the additional power circuit and permanent magnet are eliminated.

## B. The plasma plume characteristics of helix-coil multi-anode APPT

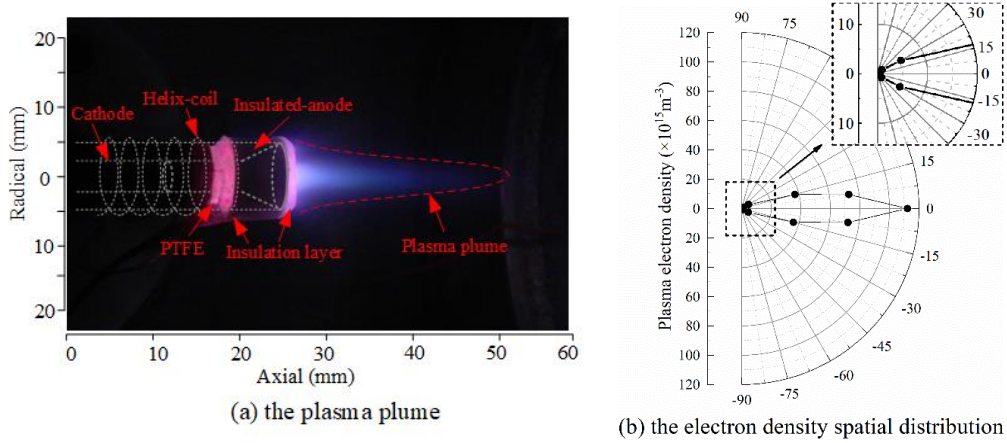

(a) the plasma plume  (b) the electron density spatial distribution
FIG. 10. The plasma plume (a) and the electron density distribution (b) of helix-coil multi-anode APPT

A series of experiments of the helix-coil multi-anode APPT have been carried out. It has been found that the waveform of interelectrode voltage and discharge current remains approximately unchanged. The amplitude of cathode current decreases to about 90% of the multi-anode APPT with the same discharge parameters. The photograph of plasma plume and the electron density distribution of helix-coil multi-anode APPT are shown in Fig. 10. It has been seen from Fig. 10(a) that the light intensity is increased and the plasma plume is further enhanced. Fig. 10(b) shows that the electron density at 0° increases to $11.4 \times 10^{16}$ m$^{-3}$, increasing by 20% compared with the multi-anode APPT. The electron density at 15° decreases to 32.3% of the density at 0°, about the same with that of the multi-anode APPT. While the electron density at 30° reduces to 4.6%, which is far below the electron density of the multi-anode APPT.

Therefore, it is concluded that the design of helix-coil is able to further reduce the diffusion of plasma and increase the density of plasma in axial direction. The improvements of plasma plume and the decrease of discharge current are supposed to be beneficial to the promotion of performance of the thruster.

## V. PERFORMANCE OF THE MULTI-ANODE APPT AND THE HELIX-COIL MULTI-ANODE APPT

To determine the experimental performance of the multi-anode APPT and the helix-coil multi-anode APPT, a number of thrust tests are performed. The microthrust measurement system used in this experiment is proposed by Liu[39], which obtains the microthrust of the pulsed electric thruster with the Polyvinylidene Fluoride (PVDF) piezoelectric thin-film sensor. The application of piezoelectric thin-film sensor in the measurement of pulsed electric thruster has been calibrated by Wong[40] and the indirect method of measuring the force exerted on the target has been confirmed to be in good agreement with the directly measured thrust [41]. The parameters of the microthrust measurement system are shown in Fig. 11. According to the previous studies, the thrust is calculated by the following equation,

$$T = \frac{2h^2 C_f}{3 K d_{31} L^2} V_{out} \qquad (7)$$

Where $T$ is the microthrust, $h$ is thickness of the PVDF thin-film (28 μm), $L$ is the length of the PVDF thin-film (15 mm), $d_{31}$ is the piezoelectric constant of the PVDF thin-film ($3.3 \times 10^{-11}$ C/N), $K$ is the voltage amplification coefficient.

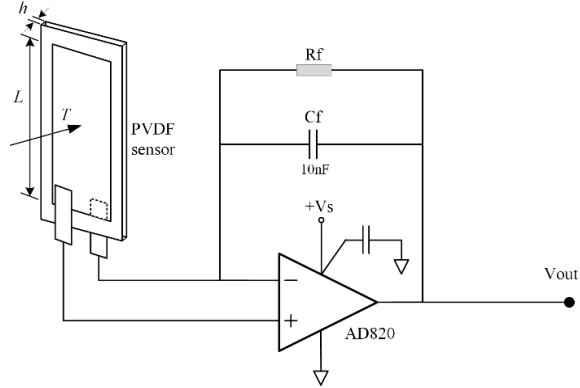

FIG. 11. Schematic diagram of the microthrust measurement system

In the test, the PVDF sensor is located at an axial distance of 120 mm from the APPT (the probe is removed before the experiment). For APPT with limited quality and volume, the impulse bit and thrust-to-power ratio are important parameters in terms of the elevation of its performance. The APPT impulse bit is obtained by the integral of the microthrust and duration time of the impulse. The energy consumed by the electrodes each impulse is calculated according to the interelectrode voltage and current waveform. The specific equations of impulse bit and thrust-to-power ratio are as follows:

$$I_{bit} = \int T dt \quad (8)$$

$$T/P = \frac{\int T dt}{\int P dt} = \frac{I_{bit}}{E} \quad (9)$$

Where $I_{bit}$ is the impulse bit, $T/P$ is the thrust-to-power ratio, $P$ is the discharge power of the electrode, $E$ is the energy consumed by the electrodes each impulse. Moreover, to investigate the variation tendency of the APPT's performance with different discharge power, the energy stored in the capacitance C is changed to alter the arc current in the discharge. The impulse bit and the thrust-to-power ratio of the APPTs as a function of energy stored in capacitance C are shown in Fig. 12.

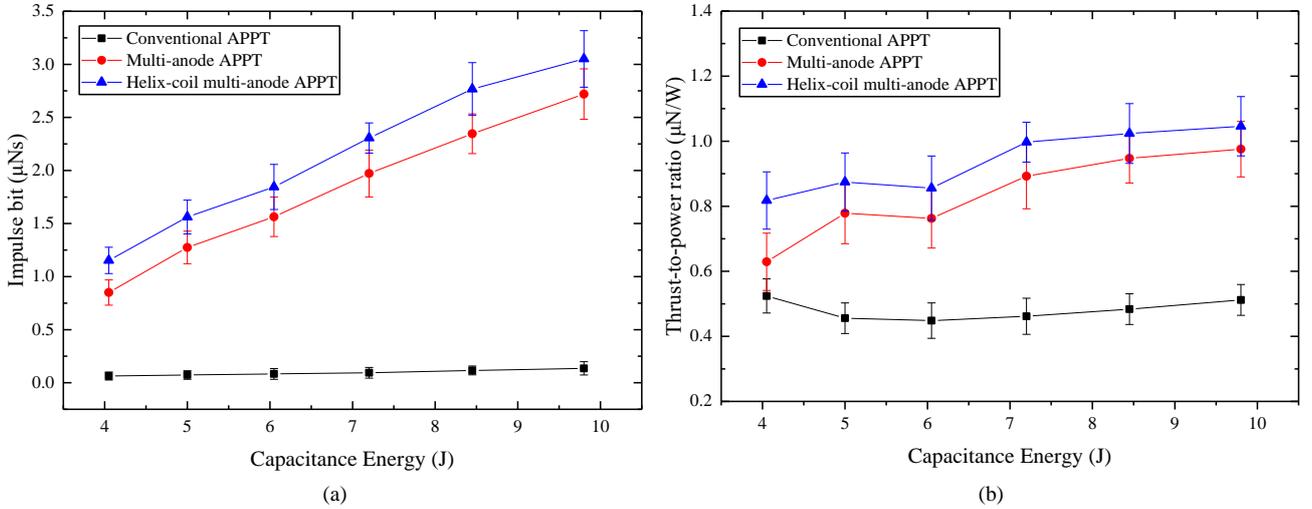

FIG. 12. The change of impulse bit (a) and the thrust-to-power ratio (b) as a function of capacitance energy.

It has been seen from Fig. 12(a) that the impulse bit of the multi-anode APPT and the helix-coil multi-anode APPT is significantly increased compared with the conventional APPT. The promotion of impulse bit of the multi-anode APPT and the helix-coil multi-anode APPT is more obvious than that of the conventional APPT with the increase of discharge power. The maximum impulse bit of the conventional APPT in this study is below 0.2 μNs. However, the minimum impulse bit of the multi-anode APPT is more than four times of the maximum impulse bit of the conventional APPT. With the increase of discharge power, the impulse bit of the multi-anode APPT increases from 0.85 μNs at capacitance energy of 4 J to 2.7 μNs at 10 J. The impulse bit of the helix-coil multi-anode APPT has been further promoted, the maximum impulse bit reaches 3 μNs. The reason for the promotion of impulse bit for the multi-anode APPT and the helix-coil multi-anode APPT at the same energy of C is concluded to be the increase of axial plasma density. The discharge current will increase synchronously with the increase of the energy stored in C. Therefore, the pinch effect of multi-anode APPT and the helix-coil multi-anode APPT has also been promoted. Combined with the increase of generation of plasma, the density of plasma in the axial direction is significantly increased. Therefore, the increase of impulse bit of the multi-anode APPT and helix-coil multi-anode APPT is more remarkable than that of the conventional APPT.

It has been found from Fig. 12(b) that the thrust-to-power ratio of the multi-anode APPT and the helix-coil multi-anode APPT is also greater than that of the conventional APPT, and they exhibit different trends with the increase of discharge power. The thrust-to-power ratio of the multi-anode APPT and the helix-coil multi-anode APPT is on the rise with the increase of discharge power. While, the thrust-to-power ratio of the conventional APPT decreases before the increase and it is basically unchanged from the minimum discharge power to the maximum discharge power. Compared with the conventional APPT, the pinch effect of the multi-anode APPT will reduce the diffusion of plasma in radial direction. The loss of energy with the charged particles is also reduced and thus the thrust-to-power ratio is promoted. The θ-pinch of helix-coil multi-anode APPT further enhances this effect. Therefore, the thrust-to-power ratio at the same energy of C is promoted. When the discharge current increases, the reduction of lost energy will be enhanced because of the improved pinch effect. As a result, the thrust-to-power ratio grows with the increase of discharge power.

Therefore, the multi-anode APPT and the helix-coil multi-anode APPT has an advanced propulsion performance through Z-pinch and θ-pinch.

## VI. CONCLUSION AND PROSPECT

This paper proposes a method of multi-anode electrode geometry to realize a special surface discharge. This type of discharge has been proved to suitable for the application in the APPT. It can effectively constrict the plasma plume outside of the anode nozzle. The multi-anode APPT and helix-coil multi-anode APPT exhibit distinct advantages on the propulsion performance compared with the conventional APPT. The innovation of the paper is mainly embodied in the following aspects:

Firstly, a special surface discharge is realized with the anode being placed far away from the insulator.

Secondly, the Z-pinch on the plasma plume leaving the nozzle is realized, which effectively reduces the diffusion of the plasma.

Thirdly, the helix-coil multi-anode APPT achieves the θ-pinch with a simple structure. It increases the density of the plasma plume with its own discharge current.

Fourthly, the multi-anode APPT and the helix-coil multi-anode APPT are able to reduce the loss of discharge energy and thus improving the performance of the thruster.

This paper focuses on the optimization of the plasma plume characteristics of the multi-anode APPT, and preliminarily proves its advantages on the propulsion. The design of the anodes is the key to realize the pinch effect, and therefore the optimized anode parameters are necessary for its application. Limited to the length of the article and the current study, these issues will be discussed in the future study.

## ACKNOWLEDGEMENT

This work was supported by National Natural Science Foundation of China (No. 51577011) and the Graduate Innovation Project of Beijing Jiaotong University (No. 2016YJS147).